\documentclass[twocolumn,showpacs,preprintnumbers,amsmath,amssymb,superscriptaddress]{revtex4-1}

\usepackage{graphicx}

\begin{document}


\title{$^{62}$Ni($n, \gamma$) and $^{63}$Ni($n, \gamma$)  cross sections measured at n\_TOF/CERN}

\author{C.~Lederer}\email[corresponding author: ]{claudia.lederer@ph.ed.uk}
\altaffiliation[Present Affiliation: ]{School of Physics and Astronomy, University of Edinburgh, UK}
\affiliation{University of Vienna, Faculty of Physics, Austria}%
\affiliation{Johann-Wolfgang-Goethe Universit\"{a}t, Frankfurt, Germany}%
\author{C.~Massimi}\affiliation{Dipartimento di Fisica, Universit\`{a} di Bologna, and Sezione INFN di Bologna, Italy}%
\author{E.~Berthoumieux}\affiliation{Commissariat \`{a} l'\'{E}nergie Atomique (CEA) Saclay - Irfu, Gif-sur-Yvette, France}%
\author{N.~Colonna}\affiliation{Istituto Nazionale di Fisica Nucleare, Bari, Italy}%
\author{R.~Dressler}\affiliation{Paul Scherrer Institut, Villigen PSI, Switzerland}%
\author{C.~Guerrero}\affiliation{European Organization for Nuclear Research (CERN), Geneva, Switzerland}%
\author{F.~Gunsing}\affiliation{Commissariat \`{a} l'\'{E}nergie Atomique (CEA) Saclay - Irfu, Gif-sur-Yvette, France}%
\author{F.~K\"{a}ppeler}\affiliation{Karlsruhe Institute of Technology, Campus Nord, Institut f\"{u}r Kernphysik, Karlsruhe, Germany}%
\author{N.~Kivel}\affiliation{Paul Scherrer Institut, Villigen PSI, Switzerland}%
\author{M. Pignatari}\affiliation{Department of Physics, University of Basel, Klingelbergstrasse 82, CH-4056 Basel, Switzerland}%
% ...
%

\collaboration{The n\_TOF Collaboration (www.cern.ch/ntof)}  \noaffiliation

\date{\today}

\begin{abstract}
The cross section of the $^{62}$Ni($n,\gamma$) reaction was measured with the time-of-flight technique at the neutron time-of-flight facility n\_TOF at CERN. Capture kernels of 42 resonances were analyzed up to 200~keV neutron energy and Maxwellian averaged cross sections (MACS) from $kT=5-100$~keV were calculated. With a total uncertainty of $4.5$\%, the stellar cross section is in excellent agreement with the the KADoNiS compilation at $kT=30$~keV, while being systematically lower up to a factor of 1.6 at higher stellar temperatures. The cross section of the $^{63}$Ni($n,\gamma$) reaction was measured for the first time at n\_TOF. We determined unresolved cross sections from 10 to 270~keV with a systematic uncertainty of 17\%. These results provide fundamental constraints on $s$-process production of heavier species, especially the production of Cu in massive stars, which serve as the dominant source of Cu in the solar system.

\end{abstract}

\maketitle

\section{\label{intro}Motivation}
The astrophysical slow neutron capture process ($s$~process) in stars produces about half of the elemental abundances between Fe and Bi. 
The $s$ process is attributed to environments of neutron densities of typically $10^{6}-10^{12}$ cm$^{-3}$, resulting in neutron capture timescales of the order of years. When an unstable nucleus is produced by neutron capture, $\beta$-decays are usually faster than subsequent neutron capture, so the reaction path follows the valley of stability. The $s$ process takes place in different stellar sites. In particular, the $s$-process abundances in the solar system are made by contributions from different generations of stars, resulting in three major components, a  $main$, a $weak$ and a $strong$ component (see e.g. \cite{KGB11}). The main component dominates in the $s$ contributions between Zr and the Pb region and is mainly associated with thermally pulsing Asymptotic Giant Branch (AGB) stars of 1 to 3 $M_\odot$ with an initial metal content close to solar \cite{BGW99}. 
During the AGB phase, He burning takes place in a shell surrounding the inert C/O core of the star. Thermal pulses are caused by He shell flashes which occur because He burning cannot sustain hydrostatic equilibrium within a thin shell. As a consequence of the mixing processes and the temperature peaks induced by the thermonuclear flashes, neutrons are released in $^{13}$C($\alpha,n$) and the $^{22}$Ne($\alpha,n$) reactions, respectively \cite{GAB98}. The strong component also originates in AGB stars but with much lower metallicities than solar \cite{TGB01}. It is responsible for about half of the solar $^{208}$Pb abundances and for the full $s$ process contribution to Bi. The weak $s$~process takes place in massive stars ($>8~M_\odot$) which later explode as supernova (e.g. \cite{WHW02}), and is producing most of the $s$ abundances in the mass region between Fe and Zr \cite{Pet68,CSA74,LHT77,RBP91a,RBP91b}. In these stars, neutrons are mostly produced at the end of convective He Core burning and during the later convective  Carbon Shell burning phase via $^{22}$Ne($\alpha,n$) reactions.  \\
The reseulting $s$-process abundances, $N_s$, depend strongly on cross sections averaged over the stellar neutron spectrum. These Maxwellian Averaged Cross Sections (MACS) are defined as
\begin{equation}
 <\sigma>=\frac{2}{\sqrt{\pi}}\frac{1}{(k_BT)^{2}}\int_{0}^{\infty}\sigma(E)E\exp(-\frac{E}{k_BT})dE
\end{equation}
where $k_B$ is the Boltzmann constant, $T$ the stellar temperature and  $\sigma(E)$ the cross section as a function of energy $E$. 
The temperatures in $s$-process environments range from 0.09 to 1 GK (GigaKelvin), corresponding to $kT$ values between 5 and 90~keV. For an accurate determination of MACSs, $\sigma(E)$  should be known up to a few hundred keV. Accurate cross sections are particularly important between Fe and Zr and for the light neutron poisons. The uncertainty of a single cross section may be propagated to the abundances of the following isotopes on the $s$-process path, or over the complete $s$-process distribution in the case of neutron poisons (see e.g. \cite{MKB10}). This propagation effect is a peculiar feature of the the weak $s$ process \cite{BG85,RHH02}. \\
To have accurate $s$-process abundances $N_{s}$ derived from precise neutron capture measurements is also of great importance for $r$-process studies because solar $r$-process abundances $N_r$ are computed as residuals of the total solar abundances $N_\odot$ after subtracting $N_s$:
\begin{equation}
N_r=N_\odot-N_s
\label{abu}
\end{equation}
Since current stellar cross sections in the Fe/Ni mass region exhibit fairly large uncertainties, a campaign was started at the neutron time of flight facility n\_TOF at CERN to measure the neutron capture cross sections of all stable isotopes of Fe and Ni with improved accuracy. Additionally, the ($n,\gamma$) cross section of the long-lived radionuclide $^{63}$Ni ($ t _{1/2}$=$101.2\pm1.5$~ yr  
\cite{CZC08}) has been studied at n\_TOF \cite{LM13}. 
This paper describes the measurement and data analysis of the ($n,\gamma$) experiments on $^{62}$Ni  and $^{63}$Ni .  \\
Current data on $^{62}$Ni($n,\gamma$) include time-of-flight measurements \cite{BSE74,BS75,TTS05,ABE08} as well as activation measurements to directly determine the MACS at $kT=25$~keV \cite{NPA05,SWa08,DFK10}. Neutron capture resonances have been analyzed over a large energy range ($E_n<200$~keV) by Beer and Spencer \cite{BS75}, while there are a few other measurements investigating only the first strong $\ell=0$ resonance at 4.6~keV \cite{HBT69,AEJ72,LVK90}. Different results for this first s-wave resonance ($\ell=0$) lead to severe differences in the low neutron energy part of evaluated cross sections, listed in libraries such as ENDF/B-VII.1 \cite{endfb7}, JENDL-4.0 \cite{jendl40} and JEFF-3.1.1 \cite{JEF05}. The n\_TOF data allowed us to determine resonance capture kernels up to 200~keV neutron energy, Maxwellian averaged cross sections cross sections were determined from $kT=5$ to 100~keV with uncertainties  between 4.5 and 10.4\%.  \\
We also measured the  $^{63}$Ni($n,\gamma$) cross section above thermal neutron energies (25 meV). Results for the resonance capture kernels and Maxwellian averaged cross sections are already published in \cite{LM13}. In this paper, we present results of the unresolved capture cross section between 10 and 270~keV. 

\section{\label{measurement}Measurement}
\subsection{Facility}
The measurements were performed at the neutron time-of-flight facility n\_TOF (Phase2) at CERN. At n\_TOF, a highly intense, pulsed neutron beam is produced by spallation reactions of a pulsed 20~GeV proton beam from the CERN Proton Synchrotron on a massive lead target. The initially very energetic neutrons are moderated by a water layer surrounding the spallation target. The resulting neutron flux approximates an energy dependence proportional to $1/E_n$ and ranges from thermal (25 meV) up to few GeV. Due to the long flight path of 185~m and a pulse width of 7~ns, a high resolution in neutron energy of $\Delta E/E\approx3\times10^{-4}$ and $\Delta E/E\approx5\times10^{-3}$ can be achieved at 1~eV, and at 1~MeV, respectively \cite{GTB13}. For a detailed description of the n\_TOF facility, see Reference \cite{GTB13}. \\
The $(n,\gamma)$ reactions on $^{62}$Ni and $^{63}$Ni were studied in separated campaigns. During the  $^{63}$Ni campaign, additional data were taken again with the $^{62}$Ni sample, because $^{62}$Ni represented the most abundant impurity in the $^{63}$Ni sample. For the final $^{62}$Ni($n,\gamma$) cross section, results from both campaigns were combined. 

\subsection{\label{detset} Detection setup}
The prompt $\gamma$-ray cascade that is emitted after each neutron capture event was detected using a pair of C$_6$D$_6$ scintillation detectors where the housing was made of carbon fibre \cite{PHK03}, in order to reduce their sensitivity to neutrons to the minimum possible value. This feature is important, since neutrons scattered from the sample can be captured in the detector material and produce $\gamma$-rays which are not distinguishable from neutron capture in the sample of interest. The C$_6$D$_6$ detector system is installed perpendicular to the neutron beam and about 9 cm upstream from the capture sample. In this configuration, background due to in-beam photons, produced at the spallation target and scattered by the sample, is minimized. Additionally, angular distribution effects of $\gamma$-rays from $\ell>0$ resonances can be neglected in this position. The C$_6$D$_6$ detectors were calibrated at 0.662, 0.898, 1.836, and 4.44~MeV using standard $^{137}$Cs, $^{88}$Y and AmBe $\gamma$-sources. Calibration runs were repeated every week during the measurement to monitor the detector stability. The data acquisition system records the full pulse shape using Flash ADCs at a sampling rate of 500 MHz, corresponding to a time resolution of 2 ns. A trigger signal from the Proton Synchrotron, shortly before the proton bunch hits the neutron target, starts the data acquisition. Data are recorded for 16~ms in the  8 MBytes on-board buffer memory of the digitizers, covering  the neutron energy range down to 0.7~eV. In the second campaign, the data acquisition system was adjusted to a recording time of 80~ms, thus extending the minimum measurable neutron energy to 27~meV. 

\subsection{Samples}
The $^{62}$Ni sample consisted of 2 g metal powder, which was pressed into a stable pellet 20 mm in diameter and about 1~mm in thickness. 
The $^{63}$Ni sample was produced about 30 years ago by breeding a highly enriched $^{62}$Ni sample in the ILL high flux reactor at Grenoble \cite{HMJ92,Muthig1984,Trautmannsheimer1992}. 
A first analysis of this material confirmed that it was free of any detectable impurities apart from the ingrown Cu component. After a chemical separation of the Cu, the remaining Ni fraction was converted into NiO grains typically 1 to 2 mm in size and with a total mass of 1156 mg. Finally, the grains were sealed in a light cylindrical container made from polyether-ether-ketone ([C$_{20}$H$_{12}$O$_{3}$]$_n$, PEEK, wall thickness 0.15~mm), with a total weight of 180 mg. Mass spectroscopic analysis of the sample yielded a $^{63}$Ni to $^{62}$Ni ratio of $0.123\pm0.001$. This sample was used for measuring the $^{63}$Ni($n,\gamma$) cross section \cite{LM13} and for fitting the first large $^{62}$Ni($n,\gamma$) resonance at 4.6~keV due to its smaller thickness (see Section \ref{4keVres} for details). Additionally to the Ni samples, a Au sample of the same geometry as the Ni samples was used to normalize the cross section. A summary of the samples is shown in Table \ref{table}.

\begin{table}[!htb]
\caption{\label{table}Sample characteristics. All samples were of cylindrical shape and 2~cm in diameter. }
\begin{ruledtabular}
\begin{tabular}{llrccc}
Sample     & Mass   & \multicolumn{2}{c}{Enrichment (w\%)} &Thickness        & Chemical       \\
           &  (mg)  & $^{62}$Ni    &  $^{63}$Ni        &(10$^{-3}$ atoms/b) & form \\
\hline
$^{62}$Ni   & 1989 & 98.0 &  - & 6.20 & metal pellet              \\
$^{63}$Ni   & 1156 & 69.4 & 8.4  &5.68 & oxide grains              \\
$^{197}$Au  & 596 & - &  - &0.584 &  metal foil \\
\end{tabular}

\end{ruledtabular}
\end{table}

\section{\label{analysis}Data Analysis}
\subsection{Determination of the Capture Yield}
All time-of-flight spectra were corrected for dead-time effects, which never exceeded 1\%. The count rate $C$ measured in a capture experiment is related to the capture yield $Y_c$ via
\begin{equation}
C(E_n)=Y_c(E_n)\phi_n(E_n)\varepsilon_c+B(E_n)
\end{equation}
with $\phi_n(E_n)$ being the number of neutrons hitting the sample, $\varepsilon_c$ the detection efficiency for capture events, and $B(E_n)$ the background reactions. 
To obtain the detection efficiency which is independent of the de-excitation path of the compound nucleus, we applied the Pulse Height Weighting Technique \cite{MG67}.  Usually, the detection efficiency for a single $\gamma$-ray depends strongly on its energy, but by  subjecting a pulse height dependent weight to each recorded signal, one can achieve a detection efficiency 
\begin{equation}
\varepsilon_c=k\times E^*
\end{equation}
which is a linear function of the excitation energy $E^*$ of the compound nucleus,  regardless of the decay pattern of the capture cascade. 
The excitation energy $E^*$  is the sum of the reaction Q-value (6.84 MeV and 9.66 MeV for $^{62}$Ni and $^{63}$Ni, respectively)  and the neutron energy in the center of mass system.
The weights can be parametrized with a polynomial function of the energy deposited in the detector. Weighting functions were determined by simulating the detector response to mono-energetic $\gamma$-rays using GEANT-4 \cite{Gea03}, implementing a detailed geometry of the experimental setup. \\
After weighting, the capture yield $Y_c$ can be calculated as 
\begin{equation}
Y_c(E_n)=N\frac{C_w(E_n)-B_w(E_n)}{E^*\phi_n(E_n)}
\end{equation}
where $C_w$ is the weighted count rate, $B_w$ the weighted background, $N$ a normalization factor, and $\phi_n$ the neutron flux incident on the sample. We used a neutron flux evaluated using long term measurements with several detectors and Monte Carlo simulations \cite{BG13}. The uncertainty in the neutron flux is 2\% below 10~keV and above 100~keV, and 4-5\% between 10-100~keV neutron energy. 
To obtain the absolute capture yield, the absolute detection efficiency, and the fraction of the neutron flux incident on the sample (beam interception factor) need to be known. 
After applying weighting functions, the efficiency to detect a capture event for each isotope only depends on the excitation energy of the compound nucleus. The systematic uncertainty in the capture yield ascribed to the Pulse Height Weighting Technique is 2\% \cite{AAA04}. The normalization factor for obtaining the absolute detection efficiency $N$ is then the same for all measured isotopes after scaling the weighted counts with the excitation energy  $E^*$. The beam interception, together with the normalization factor $N$ was determined with the saturated resonance technique at the $E_n=4.9$~eV resonance in Au, using a Au sample of the same diameter as the Ni samples. If the Au sample is chosen sufficiently thick, no neutrons are transmitted through the sample at the resonance energy. Since the capture width $\Gamma_\gamma$ is bigger than the neutron width $\Gamma_n$ for this resonance,  almost all neutrons interacting with the sample get captured. It has been demonstrated in Ref. \cite{BAG07}   that a
normalization obtained from this saturated resonance in Au is nearly independent of even large changes in the resonance parameters. \\
Since the neutron beam profile changes with neutron energy, the beam interception factor depends slightly on neutron energy as well. This effect was determined by Monte Carlo simulations \cite{GTB13}. In the investigated neutron energy range the beam interception factor never changed by more than $\pm1.5$\% compared to the value at 4.9 eV. We estimated the systematic uncertainty of the final cross section due to the normalization $N$ and the beam interception, including a possible misalignment of the sample which would affect the energy dependence of the beam interception, as 1\%. The resulting total systematic uncertainty for determining the absolute capture yield  is consequently 3\% up to 10~keV and from 100-200~keV, and 5.5\% from 10-100~keV neutron energy. \\ 
The effective neutron flight path, and thus the neutron energy calibration, was determined relative to low energy resonances in Au, which have been recently measured at the time-of-flight facility GELINA with high precision \cite{MBK11}. 
 
\subsection{Backgrounds}
The background for capture measurements at n\_TOF consists of a number of different components. \\
Ambient background is coming from cosmic rays, natural radioactivity and a possible radioactivity of the sample itself. This background is determined  by runs without neutron beam.  \\

Sample-independent background, due to reactions of the neutron beam with any structure material, is determined in runs with an empty sample holder. \\

\begin{figure}[!htb]
\begin{center}
\includegraphics[width=8.2 cm]{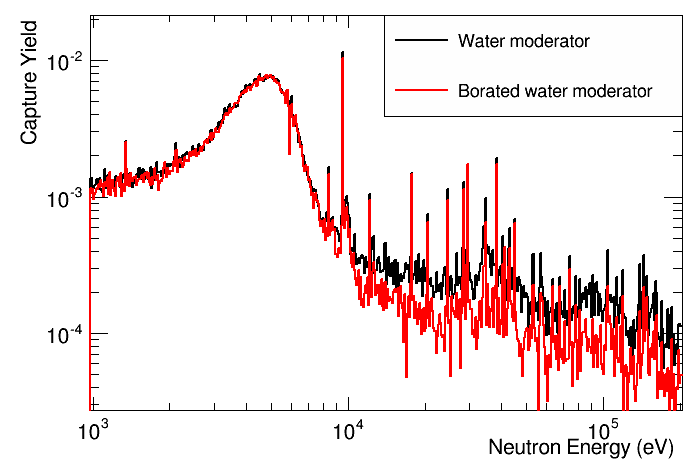}
\caption{(Color online) Capture yield of $^{62}$Ni using water (black) and borated water (red) as moderator. The addition of boron yields a significant reduction of the photon induced background in the keV region.  \label{yearcomp}}
\end{center}
\end{figure}

Sample-dependent background consists of two components. Neutrons, scattered from the sample into the experimental area where they are captured, and photons, which are produced at the spallation target and are scattered from the sample into the detector. The latter background, called in-beam $\gamma$ background, appears at neutron energies between 10 and 300~keV. It stems mainly from neutron capture on the hydrogen of the moderator and could be significantly improved in the second campaign by using borated water as moderator. This improvement is demonstrated in Fig. \ref{yearcomp}, which shows a comparison of the $^{62}$Ni capture yields from both campaigns, using water in the first, and borated water in the second campaign. \\

\begin{figure}[!htb]
\begin{center}
\includegraphics[width=8.2 cm]{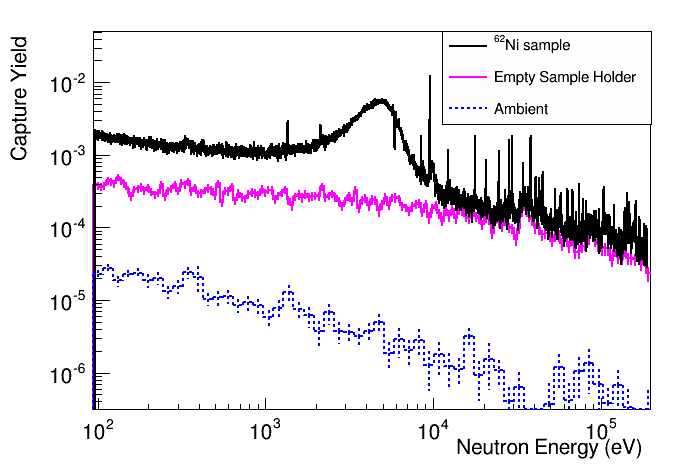}
\includegraphics[width=8.2 cm]{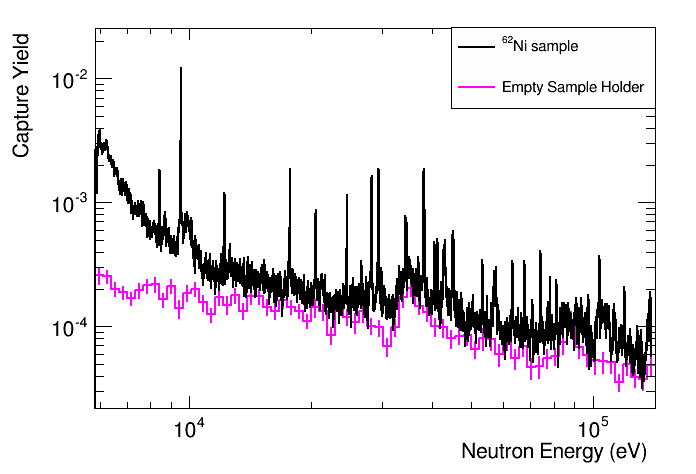}
\caption{(Color online) (Top) Capture yield of  $^{62}$Ni (black, solid line) compared with backgrounds due to neutron reactions in surrounding materials (pink, solid line, measured with empty sample holder) and ambient background (blue, shaded line). While the ambient background is 2 orders of magnitude smaller than the signal over the whole energy range, the empty background plays a crucial role in the higher keV range. (Bottom) Zoom into the neutron energy region from 6 to 100~keV. \label{ni62comp}}
\end{center}
\end{figure}

\begin{figure}[!htb]
\begin{center}
\includegraphics[width=8.2 cm]{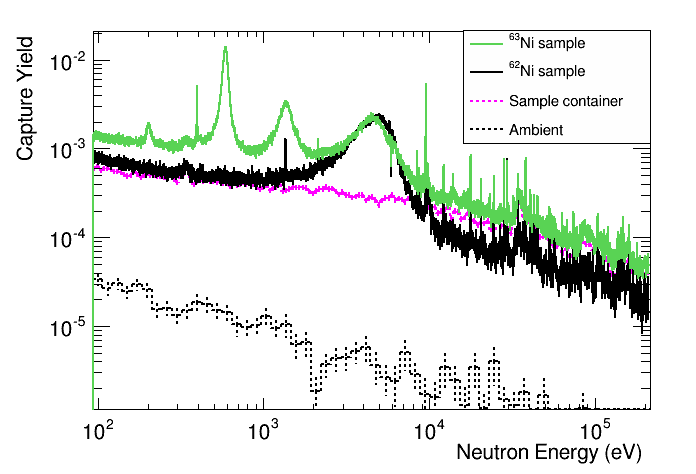}
\caption{(Color online) Capture yield of  $^{63}$Ni (green) compared with backgrounds due to neutron reactions on $^{62}$Ni in the sample (black) and with surrounding materials (pink, shaded line, measured with empty sample container), and ambient background (black, shaded line). The spectrum recorded with the $^{62}$Ni sample was scaled to the areal density of $^{62}$Ni in the $^{63}$Ni sample. \label{ni63comp}}
\end{center}
\end{figure}

Sample dependent backgrounds can be investigated using black resonance filters installed about halfway between the spallation target and the sample. These filters are sufficiently thick that the neutron spectrum is left void of neutrons at the energies of certain strong resonances. Accordingly, events in these energy windows are due to background reactions. We checked this background for neutron energies below 1~keV by comparing sample spectra with filters with the spectrum of the empty sample holder with filters and found no indication for such a sample related background. For higher neutron energies, this comparison was not possible due to lack of statistics. Since this background, however, is varying smoothly with neutron energy, it can be assumed as being constant over the width of the resonance and therefore be fitted while fitting the resonance shape. This approach could be cross checked by analyzing the $^{62}$Ni data from two different campaigns, each having different backgrounds (for the second campaign borated water was used as moderator, reducing the photon background). The capture kernels of $^{62}$Ni resonances mostly agreed within statistical uncertainties for both campaigns. For the few exceptions, the standard deviation of the two fits was used as uncertainty of the capture kernel. \\
Multiple scattering (MS) is a background that arises when a neutron is captured in the sample after it had been scattered within the sample itself. This background can be large in resonances with high scattering-to-capture ratios and depends also strongly on the sample geometry. The MS corrections are considered by the SAMMY code \cite{sammy}, which was used for analyzing the neutron resonances in $^{62}$Ni. For the unresolved cross section of $^{63}$Ni no such corrections could be applied due to the unknown scattering cross section. However, the effect is small since the $^{63}$Ni sample was relatively thin. A possible overestimation of the cross section due to this effect is included in the systematic uncertainty of the cross section.\\
A further sample related background   consists of $\gamma$-rays originating from inelastic scattering of neutrons. This background can be neglected in this measurement since the first excited state in $^{62}$Ni and the first excited state above the detector threshold of 250~keV in $^{63}$Ni are above 0.5~MeV. In both cases population of those levels was not possible on the investigated neutron energy range \cite{ensdf}.\\
The capture yields of $^{62}$Ni and $^{63}$Ni together with the ambient and sample-independent background components are shown in Figures \ref{ni62comp} and \ref{ni63comp}, respectively.

\section{\label{res62} Results on $^{62}$N${\rm i}$($n,\gamma$)}
\subsection{Resonance Analysis}
\begin{figure}[htbp]
  \centering
	  \begin{minipage}[b]{7.3 cm}
    \includegraphics[width=7.2 cm]{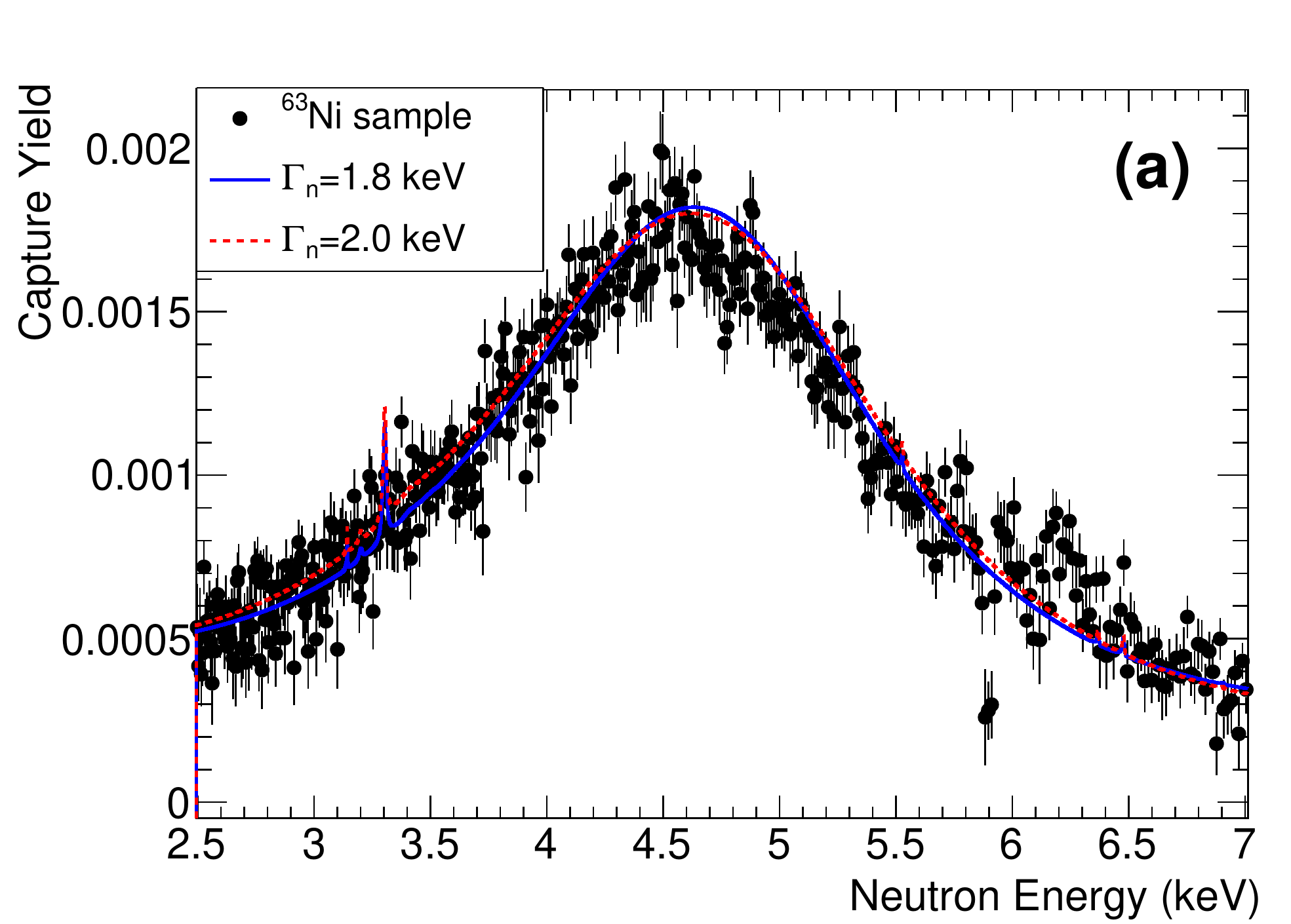}  
  \end{minipage}
  \begin{minipage}[b]{7.3 cm}
    \includegraphics[width=7.2 cm]{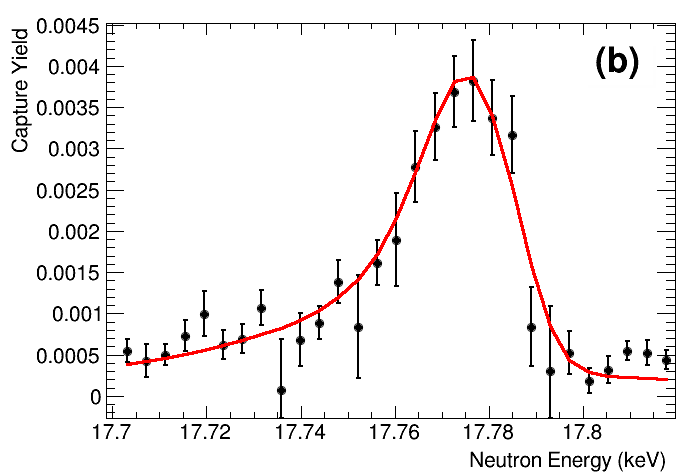}  
  \end{minipage}
  \begin{minipage}[b]{7.3 cm}
    \includegraphics[width=7.2 cm]{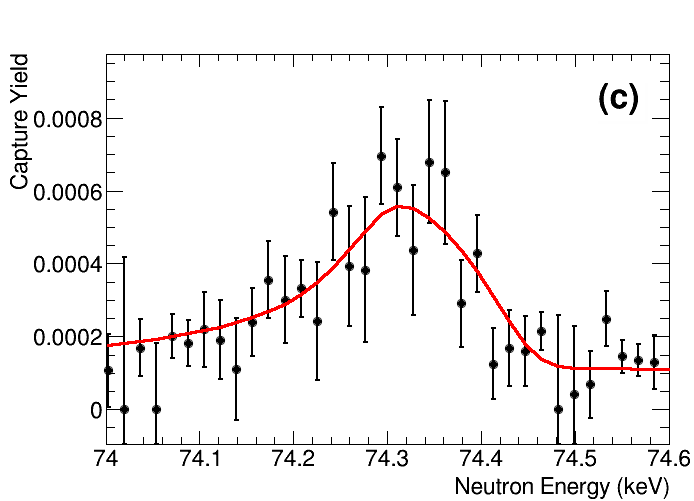}  
  \end{minipage}
	  \begin{minipage}[b]{7.3 cm}
    \includegraphics[width=7.2 cm]{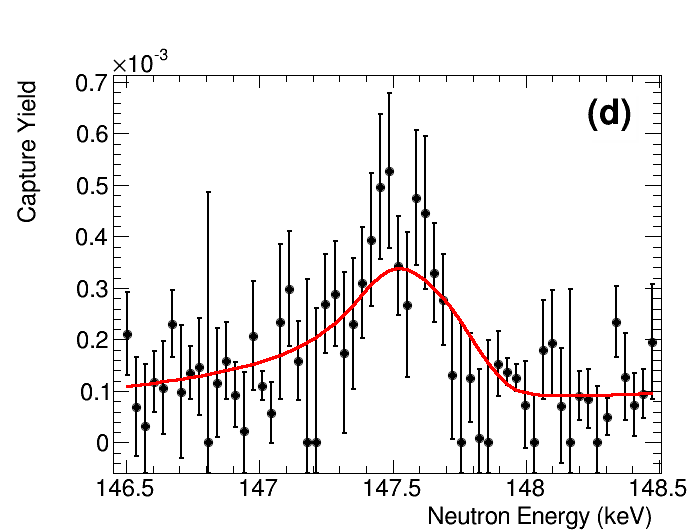}  
  \end{minipage}
  \caption{(Color online)  (a-d) Examples for resonances fitted with the program SAMMY \cite{sammy}. The dots are the measured data, the line represents the result of the resonance fit. Panel a shows the fit of the 4.6~keV resonance which was analyzed using the spectra obtained with the $^{63}$Ni sample. The data in panels b-d are from the first $^{62}$Ni campaign.  \label{res} }
  \label{nicupath}
\end{figure}

Neutron resonances up to about 200~keV neutron energy were identified and analyzed using the multi-level R-matrix Code SAMMY \cite{sammy}. 
The fitting procedure applied in SAMMY to find the 'best fit' values of parameters and the associated parameter covariance matrix is based on the Bayes' theorem. The resonance shapes were fitted using the Reich-Moore approximation, including corrections for self shielding, multiple scattering and impurities in the sample, which were mainly other Ni isotopes. Experimental effects, such as Doppler broadening and the resolution of the capture setup, were taken into account. Because the measured resonance widths were in most cases larger than the natural widths due to the broadening, only the capture kernel could be determined. It is related to the resonance area via
\begin{equation}
k_\gamma = \frac{2}{\pi\lambda^2}\int_{-\infty}^{+\infty}{ \sigma(E) dE}=g_s\frac{\Gamma_n\Gamma_\gamma}{\Gamma_n+\Gamma_\gamma}
\end{equation}
where $\lambda$ denotes the de Broglie wavelength at the resonance energy, and $\Gamma_n$, $\Gamma_\gamma$ the neutron and capture widths of the resonance. The statistical spin factor $g_s=(2J+1)/(2s+1)(2I+1)$ is determined by the resonance spin $J$, the neutron spin $s=1/2$ and the spin $I$ of the target nucleus. The results obtained from the SAMMY fits with their statistical uncertainties are listed in Table \ref{tab1} for resonances up to 200~keV.  We used the partial neutron widths $\Gamma_n$ obtained by Beer and Spencer \cite{BS75} for $\ell=0$ resonances to fit the radiative width $\Gamma_\gamma$. For resonances with $\ell>0$, no experimental data for partial widths were available, so the capture kernel $k_\gamma$ is given in the table. Examples for resonance fits are shown in Fig. \ref{res}.  Table \ref{tab1} lists the combined result and propagated statistical uncertainties of both measurement campaigns. The systematic uncertainties due to the pulse height weighting (2\%), the normalization (1\%), and the neutron flux shape (2-5\%) are not included in Table \ref{tab1}. This leads to a total systematic uncertainty in the capture kernel of 3\% for resonances up to 10~keV and from 100-200~keV, and 5.5\% for resonances from 10-100~keV.  

\begin{table}[!htb]
\caption{\label{tab1}Resonance energies $E_R$ and capture kernels $k_\gamma$ of the $^{62}$Ni($n,\gamma$) reaction. When possible,  $\Gamma_\gamma$ values have been fitted using spin assignments and  $\Gamma_n$ values from Beer and Spencer \cite{BS75}. Resonances, which were not seen in any previous measurement are marked by an asterisk. }
\begin{ruledtabular}

\begin{tabular}{ccccc}
$E_R$ (eV)     & $g_s$ & $\Gamma_n$ (meV) & $\Gamma_\gamma$ (meV)      & $k_\gamma$ (meV)        \\ \hline
$2128.6\pm0.2$ &         &            &                      & $0.570\pm0.043$   \\
$4614.8\pm6.8$ &   1       &            &   $2545\pm143$   &   \\
$8438.4\pm1.1$ &        &            &                      & $11.1\pm0.5$      \\
$9540.3\pm0.7$ &        &            &                      & $146.4\pm6.1$    \\
$12225.4\pm1.7^{*}$ &       &            &                      & $ 15.6\pm2.5$   \\
$17791.5\pm1.4$ &       &            &                      & $ 52.6\pm2.2$     \\
$20602.3\pm1.5^{*}$ &      &             &                      & $ 37.6\pm1.8$        \\
$24621.9\pm0.5$ &      &            &                      & $ 76.9\pm3.4$      \\
$28417.5\pm3.0$ &      &            &                      & $ 124.4\pm5.1$       \\
$29507.1\pm3.2$ &        &            &                      & $ 211.2\pm8.6$       \\
$29960.1\pm2.4^{*}$ &     &            &                      & $ 13.9\pm2.0$   \\  
$34473.5\pm6.4$ &    &            &                      & $ 114\pm12$        \\
$38279.5\pm1.8$ &    &            &                      & $ 313\pm17$       \\
$40547.8\pm2.2$ &    &            &                      & $ 56.8\pm6.6$    \\
$41241.6\pm2.6$ &    &            &                      & $59\pm12$       \\
$43023\pm19$ &  1 &    340000   &      $496\pm45$       &      \\
$45137.1\pm2.1$ &    &            &                      & $160\pm9$        \\
$53402.4\pm6.0^{*}$ &    &            &                      & $81\pm14$        \\
$57024\pm15$  &    &            &                      & $108\pm15$       \\
$57634\pm9^{*}$  &    &            &                      & $71\pm10$       \\
$63443.6\pm2.9$ &    &            &                      & $90\pm25$         \\
$67911.8\pm2.6^{*}$  &    &            &                      & $75\pm28$        \\
  $70892.9\pm3.2^{*}$  &    &            &                      & $ 61\pm12 $  \\
  $74419.6\pm2.6$  &    &            &                      & $186\pm15$  \\
 $77463\pm25$  & 1 &  70000 &  $265\pm53$   &   \\
  $78519.3\pm8.1$  &    &            &                      & $130\pm14$  \\
  $81469\pm31^{*}$ &    &            &                      &  $79\pm13$ \\
  $93944\pm46$  &    &            &                      &  $114\pm29$  \\
  $95038\pm1033$  & 1 &  2500000 &   $<1200$  &      \\
 $104168\pm22$  &    &            &                      &  $371\pm73$    \\
 $106550\pm1460$   & 1 &4600000 &    $<3300$   &    \\
  $113203.2\pm6.7$  &    &            &                      &  $208\pm44$    \\
$120052\pm47$  &    &            &                      &  $323\pm54$\\  
 $131919\pm15^{*}$  &    &            &                      &  $174\pm36$ \\
 $139011\pm45$  &    &            &                      &  $504\pm84$ \\
$144191\pm25$  &    &            &                      &  $488\pm135$ \\
$147713\pm32^{*}$  &    &            &                      &  $568\pm60$\\
 $149873\pm66$  & 1 &140000&    $584\pm117$   &   \\
$161745\pm19$  &    &            &                      &  $557\pm94$\\
 $170593\pm21^{*}$  &    &            &                      &  $403\pm70$ \\
$180902\pm21^{*}$  &    &            &                      &  $445\pm91$ \\
 $187175\pm45$  & 1 &  90000   &     $1610\pm296$      &  \\

\end{tabular}
\end{ruledtabular}
\end{table}

\subsubsection{Resonance at $E_R=4.6$~keV \label{4keVres}}
The shape of the neutron resonance at $E_R$=4.6~keV is highly affected by background from multiple scattering, due to its very high scattering-to-capture ratio of $\approx$ 800. It was found impossible to fit this resonance with the relatively thick $^{62}$Ni sample, therefore data measured with the thinner $^{63}$Ni sample, where multiple scattering is less important, were used to analyze this resonance.  Since the estimated multiple scattering background varies with the neutron width $\Gamma_n$, the resonance was fitted while keeping $\Gamma_n$ constant. The resonance was assigned as $\ell=0$ due to its shape. Using two previously measured values for the neutron width, $\Gamma_n=1.822$~keV \cite{LVK90} and $\Gamma_n=2.075$~keV \cite{AEJ72},  $\Gamma_\gamma$ values of 2.4~meV and 2.7~meV were obtained in the SAMMY fits, respectively. A resonance fit was not possible using a third experimental value for $\Gamma_n$ of 1.3~keV \cite{GRH71}. Since this resonance is an s wave, the two possible options for the fit yield different contributions to the cross section at lower energies. In fact, the thermal cross section obtained with the two choices is 16.2 barn for $\Gamma_\gamma=2.7$~meV, but only 12.8 barn for $\Gamma_\gamma=2.4$~meV.  Previous measurements of the thermal cross section result in reported values between 14.0 and 21~barn \cite{Pom52,MPT56,HH62,SJ70,Ish73,MNN74,IRP77,VP97}, with the majority of values grouped around 14.5~barn \cite{Pom52,HH62,SJ70,MNN74,IRP77}. Due to this large spread, these previous measurements cannot give us a hint on the correct $\Gamma_\gamma$ value. A new measurement of this resonance using a much thinner sample would be desirable in the future, especially since this resonance contributes about 50\% to the Maxwellian averaged cross section (MACS) at $kT$=30~keV. 

\subsubsection{Level Spacing}

\begin{figure}[!htb]
\includegraphics[width=9. cm]{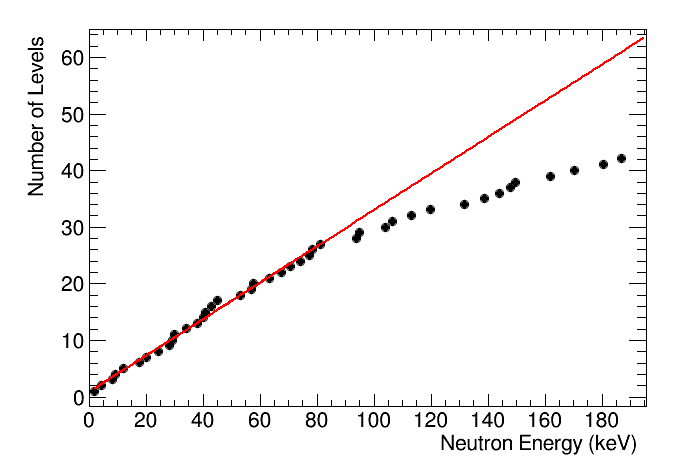}
\caption{(Color online) Accumulated number of levels as a function of neutron energy. The black dots represent the data, the red line is a linear fit from 0 to 80~keV.  \label{levels}}
\end{figure}

It is expected that the average level density of the compound nucleus is constant over the investigated energy range. Figure \ref{levels} shows that the accumulated  number of observed levels as a function of neutron energy follows the expected  linear behaviour up to about 80~keV. The increasing number of missing levels is due to the weakening signal-to-background ratio combined with the decreasing energy resolution of the n\_TOF setup. We find an average level spacing  of roughly 28~keV for s-wave resonances and 3.4~keV for $\ell>0$ resonances. 
The consequences of missing resonances for the Maxwellian averaged cross sections are discussed in section \ref{stell}.

\subsection{Maxwellian Averaged Cross Sections \label{stell}}

We calculated Maxwellian averaged cross section from $kT=5-100$~keV using the resonance parameters obtained from the SAMMY fits. Resonances parameters from 200~keV onwards were taken from the JENDL-4.0 library \cite{jendl40}. The MACS values from $kT=5-100$~keV together with their statistical and systematic uncertainties are listed in Table \ref{tab2} and Table \ref{tab3} details the uncertainties for three typical values of $kT$.  Systematic uncertainties include the Pulse Height Weighting Technique, the normalization, and the neutron flux. The impact  of the two different fits for the 4.6~keV resonance according to the different multiple scattering corrections has been included as separate systematic uncertainty (called "`MS at $E_R=4.6$~keV"' in Table \ref{tab3}). \\
To investigate the effect of missing levels on the MACS values an average cross section was calculated from our data in the energy range from 81 to 200 keV, using simulated self shielding and multiple scattering corrections.  These corrections were obtained by means of Monte Carlo simulations taking into account the sample geometry and neutron capture and scattering cross sections. The MACS values of this approach were between 3\% and 7\% higher in the range $kT=40-100$~keV than the results calculated from resonance data only. 
We included this difference as additional systematic uncertainty in Table \ref{tab3} (missing levels).

\begin{table}[!htb]
\caption{\label{tab2}Maxwellian averaged cross sections of the $^{62}$Ni($n,\gamma$) reaction from 5 to 100~keV together with statistical and systematic uncertainties. }
\begin{ruledtabular}

\begin{tabular}{lccc}
$kT$ (keV)     & MACS (mb) & \multicolumn{2}{c}{Uncertainty (\%)}   \\
\cline{3-4}
         &             & Statistical  &  Systematic       \\
\hline
~5     &   181.2     &    0.6      & 5.2           \\
10     &    83.2    &   0.6      &4.9             \\
15     &    50.8   &     0.6     & 4.8           \\
20     &    35.8   &     0.7     & 4.4            \\
25     &    27.4   &    1.0    & 4.3             \\
30     &    22.2   &     1.5     & 4.2               \\
40     &    16.0  &       2.7   & $-4.1/+5.1 $      \\
50     &    12.5  &       3.8    & $-4.1/+6.7 $       \\
60     &    10.2   &      4.7    & $-4.0/+7.2$        \\
80     &    7.44 &       6.0    & $-3.9/+8.0 $   \\
100     &   5.75 &        6.7    &$-3.8/+8.0  $ \\
\end{tabular}
\end{ruledtabular}
\end{table}

\begin{table}[!htb]
\caption{\label{tab3}Contributions to the uncertainties (in \%) for the stellar $^{62}$Ni($n,\gamma$) cross sections (see text for details).}
\begin{ruledtabular}
\begin{tabular}{lccc}
$kT$  (keV)         & 5     & 30            & 100 \\ \hline
Weighting Functions  & 2   & 2          &2 \\
Normalization        & 1   & 1           & 1 \\
Neutron Flux Shape  & 2.0 & 2.7         & 2.9\\
MS at $E_R=4.6$~keV & 4.2 &2.3          & 0.9 \\
Missing Levels          & -     & -          & +7 \\
Counting Statistics & 0.6 & 1.5         & 6.7 \\ \hline
Total               & 5.2 & 4.5         & -7.7/+10.4 \\

\end{tabular}
\end{ruledtabular}
\end{table}

\begin{figure}[!htb]
\includegraphics[width=9. cm]{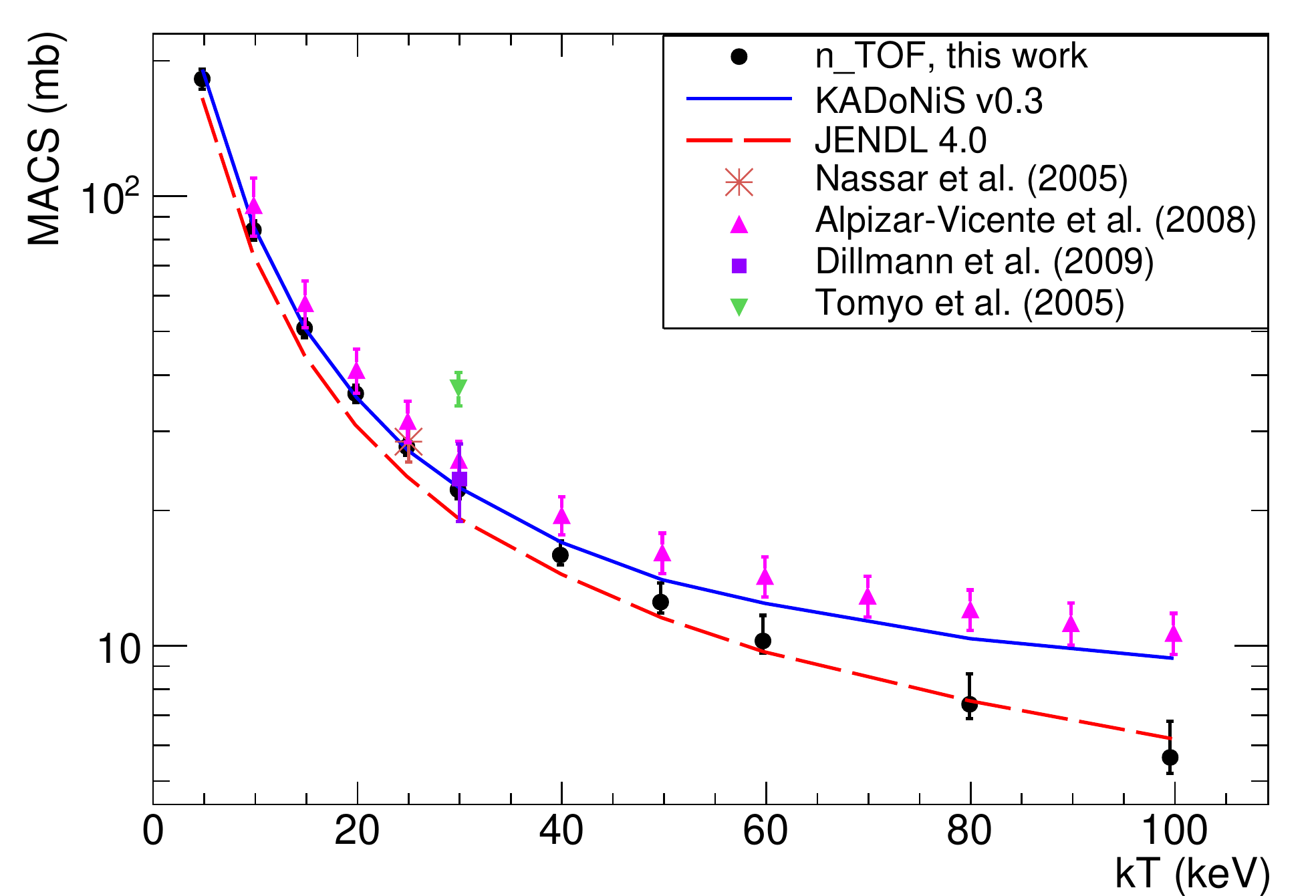}
\caption{(Color online) Maxwellian Averaged Cross sections from 5 to 100~keV compared to previous measurements (Alpizar-Vicente et al. \cite{ABE08}, Nassar et al. \cite{NPA05}, Dillmann et al. \cite{DFK10} and Tomyo et al. \cite{TTS05}). The results obtained with data from the JENDL-4.0 evaluation (dashed line, \cite{jendl40}) and the recommended MACS values of the KADoNiS compilation  (solid line, \cite{DRK09}) are included as well. \label{macs}}
\end{figure}

A comparison of our results to previous measurements and evaluations is shown in Fig. \ref{macs}. For $kT<30$~keV, the n\_TOF data are in agreement with the results of Alpizar-Vicente et al. \cite{ABE08}. At 25~keV and 30~keV, our MACS is in excellent agreement with activation measurements of Nassar et al. \cite{NPA05} and Dillmann et al. \cite{DFK10}, while being significantly lower than a previous time-of-flight measurement by Tomyo et al. \cite{TTS05}. Towards higher $kT$ values, our data start to deviate from the results of Alpizar-Vicente et al. \cite{ABE08}, being systematically lower up to a factor of 1.8. As investigated by Monte Carlo simulations, missing levels due to high background at high neutron energies cannot account for that difference.  For $kT>50$, our data are in fair agreement with MACS calculated using resonance parameters of the JENDL-4.0 evaluation \cite{jendl40}, which is mainly based on a measurement by Beer and Spencer \cite{BS75}.

\section{\label{res63} Results for $^{63}$N${\rm i}$($n,\gamma$)}
The resonance analysis for the $^{63}$Ni($n,\gamma$) reaction has already been described and published in Ref. \cite{LM13}. In this section, the results for the unresolved cross section of $^{63}$Ni($n,\gamma$) are presented. From 10~keV onwards, we calculated an averaged cross section, since the high background, mainly due to reactions of neutrons with $^{62}$Ni and with the sample container prevented us from analyzing more resonances. The $^{63}$Ni($n,\gamma$) capture yield was calculated by subtracting the background due to $^{62}$Ni($n,\gamma$) reactions using the spectra recorded with the $^{62}$Ni sample and the known $^{62}$Ni abundance in the $^{63}$Ni sample.  Background from reactions on oxygen is negligible, due to the low reaction cross section. 
The average cross section was calculated using the thin target approximation
\begin{equation}
\sigma=\frac{Y_c}{n},
\end{equation}
where $n$ is the areal density of the sample and $Y_c$ the neutron capture yield. As for $^{62}$Ni, systematic uncertainties are coming from the Pulse Height Weighting Technique (2\%), the neutron flux (2\%-5\%), and the normalization (1\%). Additionally, the $^{62}$Ni/$^{63}$Ni ratio in the sample contributed an uncertainty of $\leq 2$\%. The background subtraction due to reactions on $^{62}$Ni in the sample introduces the largest systematic uncertainty, which we estimated as 15\% based on different ways to treat the background at $^{62}$Ni resonances. Assuming a high multiple scattering correction of 5\%, the total systematic uncertainty of this measurement amounts to 17\%. The cross sections from 10-270~keV, along with statistical uncertainties, are listed in Table \ref{tab4}. MACS values and the implications of the $^{63}$Ni cross section for stellar nucleosynthesis have been published in \cite{LM13}.

\begin{table}[!htb]
\caption{\label{tab4} Average $^{63}$Ni($n,\gamma$) cross sections between 10 and 270~keV laboratory neutron energy with statistical uncertainties. The total systematic uncertainty is 17\%.}
\begin{ruledtabular}

\begin{tabular}{ccc}
 \multicolumn{2}{c}{Neutron Energy (eV)} & Cross Section (mb)  \\
$E_{low}$ & $E_{high}$ &    \\ \hline
    10104 & 12136  &   $87\pm3$   \\
    12136 &14577  &   $142\pm29$  \\
    14577 & 17506  &   $160\pm26$ \\
    17506 & 21023  &   $111\pm25$  \\
    21023 & 30304  &   $77\pm15$  \\
   30304 &43664  &   $57\pm22$ \\
   43664 & 62871 &   $50\pm12$   \\
	 62871 &90456 &   $37\pm8$   \\	
   90456 &130027 &   $22\pm7$   \\
	 130027 & 186705 &   $18\pm8$   \\
	186705 & 267743 &   $7.0\pm5.8$   \\
\end{tabular}
\end{ruledtabular}
\end{table}

\section{\label{impl} Astrophysical Implications}
In addition to the cross sections of the target nuclei in their ground states, as measured here, reactions on thermally excited states have to be considered in the determination of stellar reaction rates to be used in astrophysical models. For $^{62}$Ni($n,\gamma$), the population of excited states is negligible across the full energy range of $s$~process temperatures. Thus, the measured laboratory cross sections directly allow to derive the stellar rates. Due to the higher nuclear level density of $^{63}$Ni, only a fraction of the stellar rate can be constrained by a measurement of $^{63}$Ni($n,\gamma$) cross sections and theoretical corrections have to be applied as described in \cite{LM13}. \\
The impact of our new results on $^{62}$Ni($n,\gamma$) and $^{63}$Ni($n,\gamma$) on the weak $s$~process in massive stars was investigated for a full stellar model for a 25~$M_\odot$ star with an initial metal content of Z=0.02 \cite{PHH13}. The complete nucleosynthesis was followed with the post-processing NuGrid code MPPNP \cite{PH12}. 
\begin{figure}[!htb]
\includegraphics[width=9. cm]{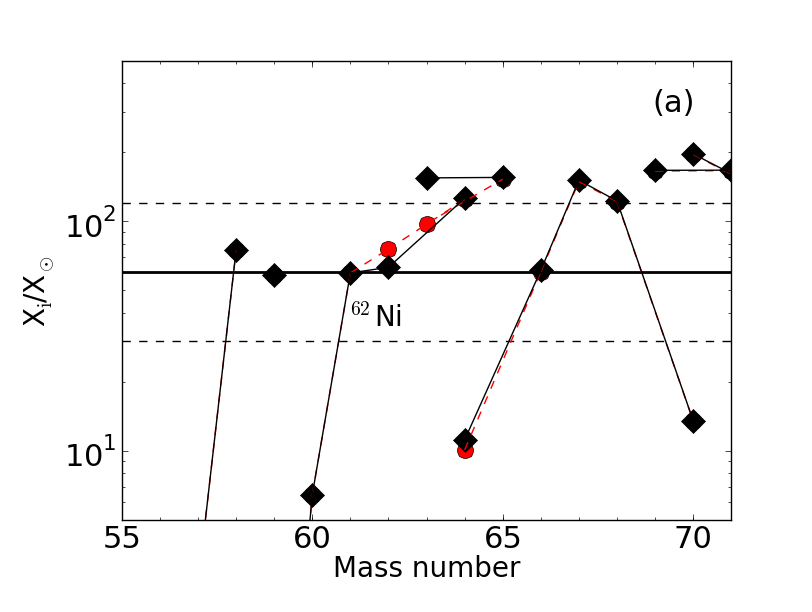}
\includegraphics[width=9. cm]{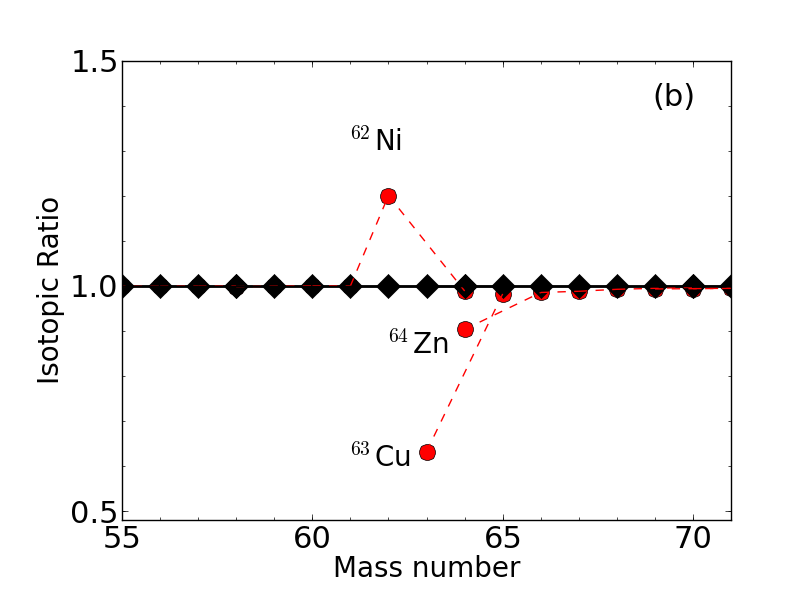}
\caption{(Color online)  (Top) Final isotopic $s$~process abundances between Fe and Ga normalized to solar system abundances. The red circles represent the abundances using the $^{62}$Ni($n,\gamma$) MACS of this work and the $^{63}$Ni($n,\gamma$) MACS reported in Ref. \cite{LM13}. This distribution is compared to the results using the recommended MACS of $^{62}$Ni and $^{63}$Ni of the KADoNiS compilation \cite{DRK09}. As a reference, the overabundance of $^{16}$O is shown as black continuous line, divided and multiplied by 2 (black dashed lines).  (Bottom) Ratio between the abundances using the new cross sections and the abundances using KADoNiS cross sections. \label{fig:abu}}
\end{figure}
Figure \ref{fig:abu} shows  the $s$-process abundance distribution in the mass region from Fe to Ga after the convective core He and the convective C shell burning phase. Although the solar system $s$-process abundances in the Ni-Cu-Zn region may be partially affected by the following core-collapse supernova event (e.g. \cite{RHH02,PGH10}), the pre-explosive $s$-process distribution is relevant as it serves as seed for the later explosive nucleosynthesis. The abundance distribution calculated with the MACSs of $^{62}$Ni and $^{63}$Ni from this work and Ref. \cite{LM13} is compared to the abundances calculated with the recommended MACS data of the KADoNiS compilation v0.3 \cite{DRK09}. Because the $^{62}$Ni MACS of this work is smaller than the value in KADoNiS for $kT>50$~keV, neutron capture rates of $^{62}$Ni in the C shell burning phase, where temperatures correspond to $kT\gtrsim90$~keV, are smaller and the final abundance of $^{62}$Ni increases by 20\%. The corresponding lower production of $^{63}$Ni results in lower abundances of $^{63}$Cu and $^{64}$Zn. This decrease is compensated for $^{65}$Cu and above $^{66}$Zn by the fact that the $^{63}$Ni MACS itself is a factor of 2 higher than the MACS value in KADoNiS, resulting in a stellar rate that is about 40\% larger than the KADoNiS rate at typical shell C burning temperatures after considering the contribution from neutron capture on excited states in $^{63}$Ni \cite{Rau12}. 
Accordingly, the $s$~abundances in this region differ only by 1-2\%. For $^{63}$Cu, which is mainly produced by the radiogenic decay of $^{63}$Ni after C shell burning, the effect of a smaller $^{62}$Ni MACS and a higher $^{63}$Ni MACS causes a 40\% lower abundance of $^{63}$Cu. Because the $^{65}$Cu abundance remains essentially unchanged, the isotopic ratio $^{63}$Cu/$^{65}$Cu is reduced after C shell burning. These results  will allow to better define the following explosive
contribution to the copper inventory of the solar system.

\section{\label{summ} Summary }
We measured the cross sections of the $^{62}$Ni($n,\gamma$) and $^{63}$Ni($n,\gamma$) reactions at the neutron time-of-flight facility n\_TOF at CERN. \\
For $^{62}$Ni($n,\gamma$), the resonance analysis was performed up to 200~keV neutron energy. In total, 42 levels could be identified, of which 12 were not seen in previous experiments. 
The Maxwellian averaged cross sections, calculated from resonance parameters is in good agreement with previous measurements for $kT$ values up to 50~keV. At higher energies our results are systematically lower than the data by Alpizar-Vicente et al. \cite{ABE08}, but in good agreement with the evaluations in the data libraries JENDL \cite{jendl40} and ENDF/B-VII \cite{endfb7}, which are mainly based on a measurement by Beer and Spencer  \cite{BS75}. Our MACS at 100~keV is also a factor of 1.6 lower than the currently recommended value of the KADoNiS compilation. \\
For the $^{63}$Ni($n,\gamma$) reaction,  the neutron resonance analysis together with the stellar cross sections are published elsewhere \cite{LM13}. We determined averaged cross sections between 10 and 270~keV with systematic uncertainties of 17\%. \\
The impact of the new stellar $(n,\gamma$) cross sections of $^{62}$Ni and $^{63}$Ni has been studied with a stellar model for a 25 $M_\odot$ star with Z=0.02. We find significant changes in the $s$~abundances of $^{62}$Ni (+20\%) and  $^{63}$Cu (-40\%), whereas the changes for heavier $s$~process isotopes are less than 2\%. These results are particularly important to understand the solar system abundances of Cu, which is dominantly produced in massive stars.   \\
 
\begin{acknowledgments}
The authors would like to thank H. Danninger and C. Gierl of the Technical University of Vienna for their help preparing the $^{62}$Ni sample. 
This work was partly supported by the Austrian Science Fund (FWF), projects P20434 and I428 and by the Federal Ministry of Education and Research of Germany, project 05P12RFFN6. M.P. acknowledges support from the Ambizione grant of the SNSF (Switzerland), from NuGrid thanks to the EU MIRG-CT-2006-046520, from the NSF grants PHY 02-16783 and PHY 09-22648 (Joint Institute for Nuclear Astrophysics, JINA), and from EuroGenesis (MASCHE). T.R. acknowledges the Swiss NSF, the European Research Council, and the THEXO Collaboration within the 7$^{th}$ Framework Program ENSAR of the EU. 
\end{acknowledgments}

\bibliography{literature}

\end{document}